\documentclass{article}
\pdfoutput=1
\usepackage[T1]{fontenc}
\usepackage[latin9]{inputenc}
\usepackage{amsmath}
\usepackage{amssymb}
\usepackage{graphicx}
\usepackage{CJK}
\usepackage{amsmath}
\usepackage{cases}
\usepackage{indentfirst}
\usepackage{cite}
\usepackage{bm}
\usepackage{subfigure}

\begin{document}
\title{Statistical Mechanics of the L-Distance Minimal Dominating Set problem}
\author{$Yusupjan\quad Habibulla$ \\School of Physics and Technology, xinjiang university, Sheng-Li Road 666,\\  Urumqi 830046}

\maketitle
\begin{abstract}
Statistical mechanics is widely applied to solve hard optimization problem, the optimal strategy related to ground state energy that depends on low temperature. Common thermodynamic process is expected to approach the ground state energy if the temperature is lowered appropriately, but this belief is not always justified when the network contains more long loops in low temperature. Previously we always implement the canonical equilibrium process to predict the low-energy, but it doesn't work in L-distance (L>1) minimal dominating set problem, because the thermodynamical process can not guarantee to find the stable state of the system at the low temperature. Here, we employ energy-clamping strategy of cavity method ( micro canonical equilibrium process ) to predict low-energy and discover that the microcanonical process still find the stable state of given system at low temperature where canonical process work out. We develop Belief Propagation Decimation (BPD) and Greedy algorithm to calculate the L-distance ($2<L<7$) minimal dominating set, we find that the BPD algorithm results outperform the Greedy algorithm. We have witnessed the emergence of negative $\beta$ with different mean energy on different L-distance. The free energy has a discontinuous phase transition at $\beta = 0$. We predict the ground state energy by microcanonical cavity method, overcoming the limitation of canonical cavity method.\\\\
\textbf{\large Keywords: }L-distance minimal dominating set problem, belief propagation, Random graph, Microcanonical ensemble, belief propagation decimation.
\end{abstract}
\section{Introduction}
Absolute temperature $T$ is one of the central concepts of the statistical mechanics, there is nothing colder than $T=0$. But, according to the thermodynamic definition of the temperature\cite{1987khuang} $1/T=\partial S/\partial E$, we can get negative temperature in high energy level. In the positive absolute temperature range the entropy increase with energy, and in the negative temperature range the entropy decrease with the energy. The negative absolute temperature demand an upper bound in energy\cite{RNF1956,KMJ1956}. The temperature is discontinuous at maximum entropy, jumping from positive to negative infinity, and negative temperature is hotter than positive temperature. If the positive temperature system and negative temperature system in the thermal contact, then the heat flow from the negative to the positive temperature system, this rise Carnot engines with an efficiency greater than unity\cite{RAMSRA2010}. Negative absolute temperature is also interest in dark energy in cosmology, where negative pressure (caused by negative temperature) is required to account for the accelerating expansion of the universe\cite{FJATMSHD2008}.\\
Statistical physics associates the probability of visiting low-energy states with low temperatures. This intuitive belief
underlies most physics-inspired strategies for addressing hard optimization problems. For example, the most popular simulated annealing (SA) dynamics, which sample low-energy configurations while gradually decreasing the temperature T,
to progress towards equilibrium configurations close to the ground states\cite{KSGCDVMP1983}. An implicit assumption in the SA is that the entropy function $S(\beta)$ is a monotonic concave function of the $\beta$ so that higher $\beta$ exclusively corresponds lower energy $E$. For the discrete-state systems, the entropy function is not always concave but is characterized by an inflection point that separates the concave higher-energy branch from the convex lower-energy branch.  The convex section of $S(E)$ also violates results obtained for canonical ensembles, suggesting a discrepancy between canonical and microcanonical ensemble analyses, as in\cite{TH2015,CADTRS2009,DM2008}. In this work we show that the entropy is still a concave function of energy for discrete-state systems, but the canonical Belief Propagation (BP) process can not always reach the ground state energy of l-distance MDS problem while mean field theory gives violates results, the symmetry of the solution space is broken when BP equation can not converge. \\
Consider simple network W formed by N nodes and M undirected links, where each link connects two different nodes. There is one set $\gamma $. If any node of the network belongs to this set, or has at least one covered neighbor node
at a distance at most $L$ that belongs to $\gamma$, then this set is called the l-distance minimal dominating set (MDS) of the given network W. Where the distance between two vertices in the graph is the minimum number of hops necessary to go from one node to the other. Each node will then investigate or monitor those nodes within a distance $L$. As an example we show in Fig. 1 a 2 and 3-distance MDS of a small network. \\
Our work inspired by the work of Haijun Zhou et al\cite{XYZYCHZHJSD2018,ZHJPRL2019}, they develop clamp-energy strategy (microcanonical ensemble process) to study the strong defensive alliance problem and obtain superior solution. We have develop statistical physics model to constructed 2-distance MDS, but we can not exactly predict the minimal state energy when $L\geq 2$. We have applied the cavity method of spin glasses\cite{MMMA2009,KFRFBJLHA2001,XJQZHJ2011,ZHJWC2012} to the 2-distance MDS problem\cite{HYQSM2019,HY2019}, we found that the cavity equation doesn't always converge at ground state. In this work we apply the energy-clamp strategy to successfully predict the minimal energy of L-distance MDS problem on Regular Random graph.\\  
\begin{figure}[!htbp]
\centering
\subfigure{
\includegraphics[width=6cm,height=4cm]{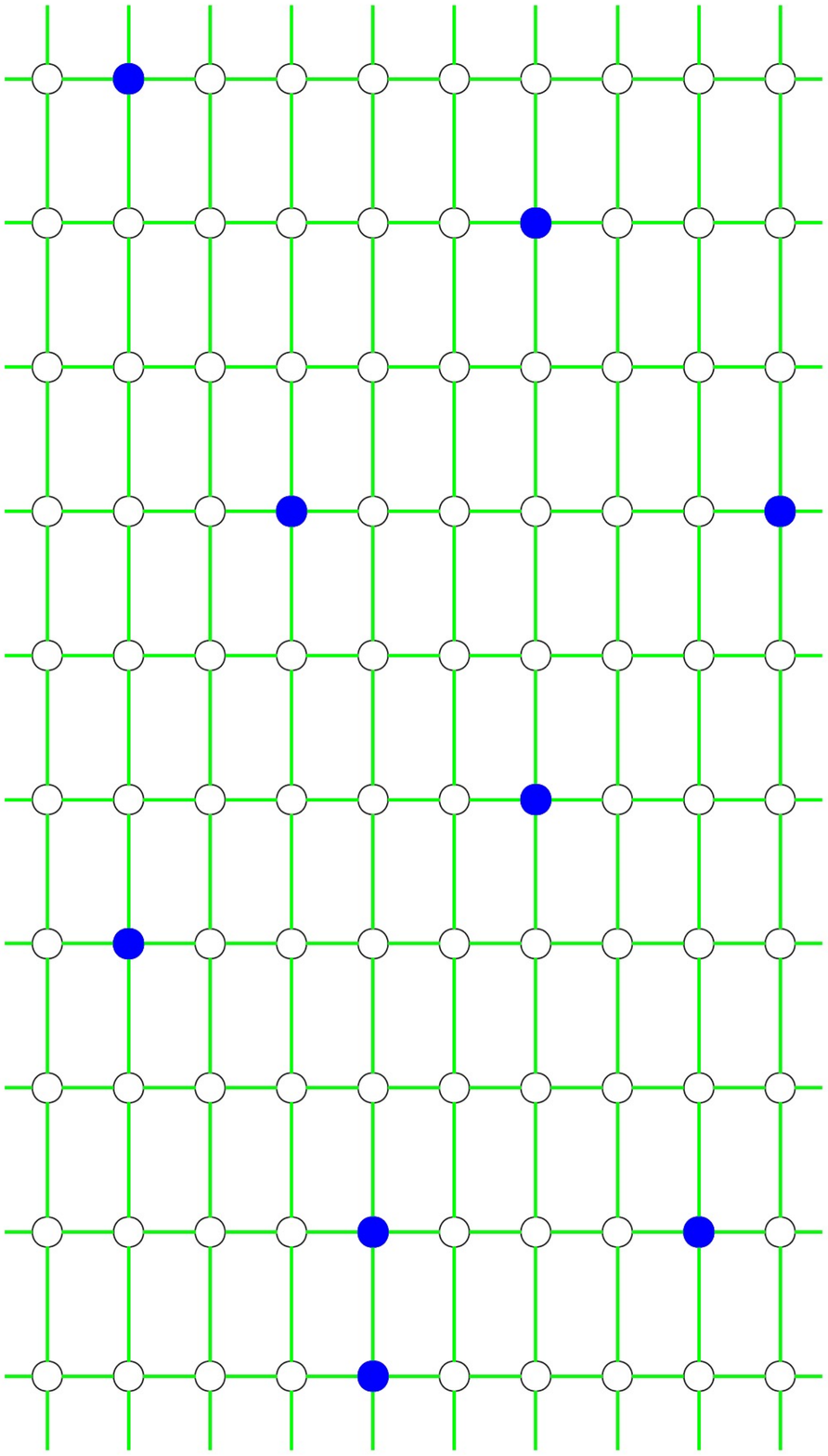}
\includegraphics[width=6cm,height=4cm]{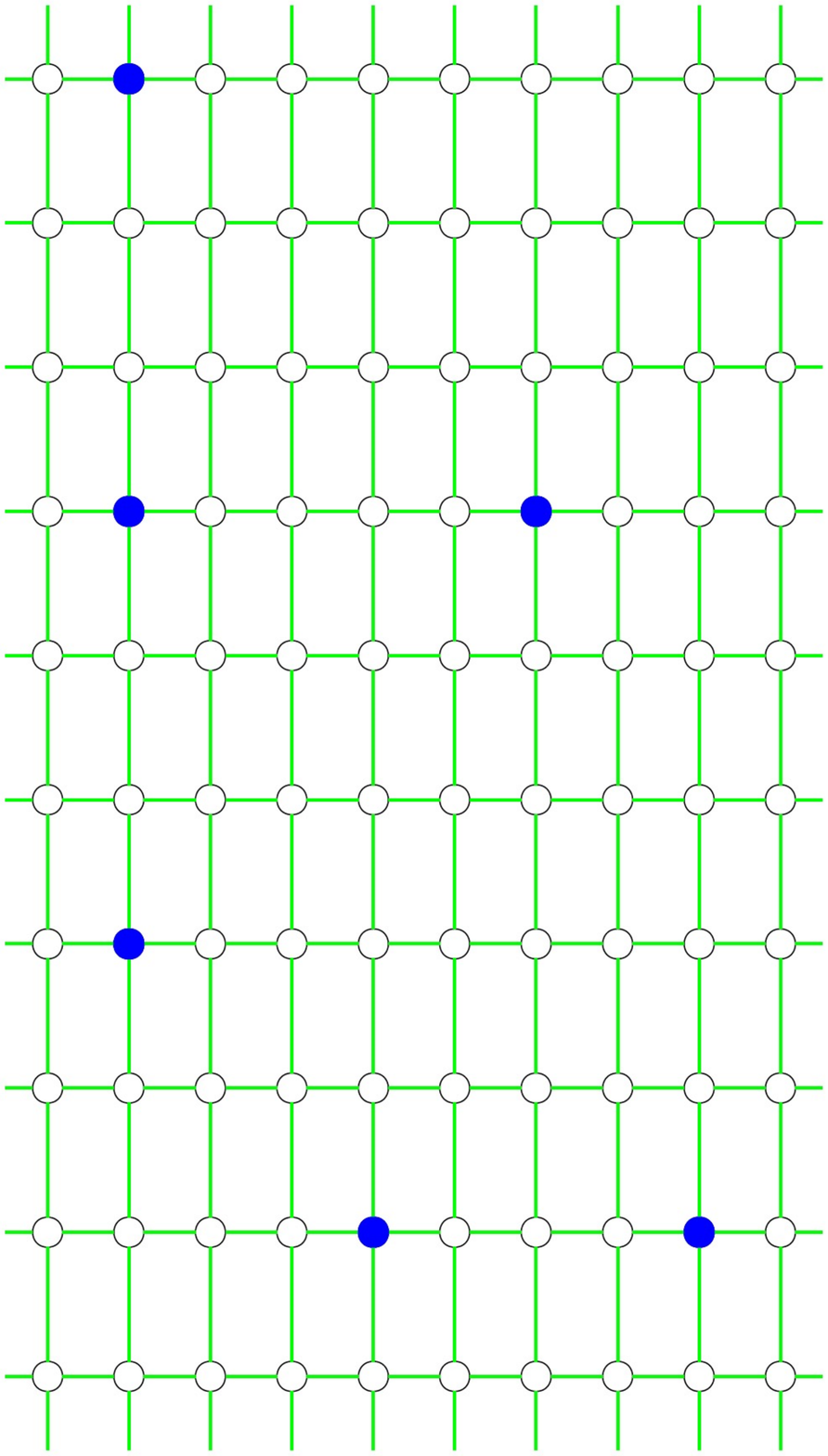}
}
\caption{(left) a solution of 2-distance minimum dominating set. (right) a solution of 3-distance minimum dominating set. a small network with N = 100 nodes and M = 200 links. Blue filled circle indicates a node being occupied, while empty circle indicates a node being empty but observed. These occupied nodes form a solution of 2 and 3-distance MDS for this network.}
\end{figure}
Recently, we have used statistical physics to study the regular minimal dominating set problem\cite{ZJHHYZHJ2015,HYZJHZHJ2015,HYJSM2017,HYCITP2018}.  We introduced belief propagation decimation (BPD), warning propagation, and survey propagation decimation algorithms to obtain the minimal dominating set. We found that our algorithms were very close to the optimal solution and the speed was very fast. The solution space has condensation transition and cluster transition on the undirected regular random (RR) graph. In the last year, we continue to use statistical physics to study the 2-distance minimal dominating set (LDMDS). The ground-state energy appears when the entropy is equal to zero on the undirected RR random graph when the mean degree changes from 3 to 9 by canonical process. We used three algorithms, population dynamics, BPD, and the greedy heuristic algorithm, to calculate the 2-distance MDS. We found that the population dynamics and BPD results were always better than those of the greedy heuristic algorithm on the single ER and RR random graphs. In this work we use clamp-energy strategy (microcanonical ensemble process) to study the L-distance MDS when $L\leq 3$, we find that the clamp-energy strategy still find ground state energy while canonical ensemble process gives violated results. \\
This paper is organized as follows: In Section 2, we introduce replica symmetry (RS) theory for the l-distance MDS problem and present the belief propagation (BP) equation and the corresponding thermodynamic quantities. In Section 3, we introduce the BPD algorithm and greedy algorithm for the l-distance MDS problem, and derive the BP equation and marginal probability equation for the different vertex state conditions. We also construct the proper BPD process to estimate the l-distance MDS. In Section 4, we introduce the energy-clamp strategy (microcanonical ensemble process) for the l-distance MDS problem. In Section 5, we draw conclusions and summarize our results. 
\section{Replica symmetry}
In this section, we introduce mean field theory for the l-distance MDS problem. We put a interaction on each vertex so that each nodes satisfies the condition of L-distance MDS, there is no external-field. In the canonical ensemble average process the partition function sum over all configuration weights at same $N,V,T$. Depending on the RS mean field theory of statistical physics, we can write partition function Z as
\begin{equation}
Z=\sum_{\underline{c}}\prod_{i\in W}{e^{-\beta \delta_{ c_{i}}^{0}}}\{1-(1-\delta_{ c_{i}}^{0})[1- \delta(c_{i}-1, min\{c_{j}\}_{j\in\partial i})]\},
\end{equation}
where the Kronecker symbol $\delta_{m}^{n}=1$ if $m = n$ and $\delta_{m}^{n}=0$ otherwise. The symbol $\underline{ c}\equiv( c_{1}, c_{2},......, c_{n})$ denotes one of the $(L+1)^{n}$ configurations, $ c_{i}=0$ if node $i$ is covered, $ c_{i}\geq 1$ if node $i$ is not be covered. If node $i$ is in state $c_{i}=0$, then it requests that the neighbor nodes only take state $c_{k}=0$ or $c_{k}=1$.  If node $i$ is in state $c_{i}=1$, then it requests that the neighbor nodes only take state $c_{k}=0$, $c_{k}=1$, or $c_{k}=2$, but at least one neighbor must be occupied. If node $i$ is in state $c_{i}=2$, then it requests that the neighbor nodes only take state $c_{k}=1$, $c_{k}=2$ or $c_{k}=3$, but at least one 2-distance neighbor must be occupied. If node $i$ is in state $c_{i}=3$, then it requests that the neighbor nodes only take state $c_{k}=2$, $c_{k}=3$ or $c_{k}=4$, but at least one 3-distance neighbor must be occupied so that at least one $c_{k}=2$ and so on. If node $i$ is in state $c_{i}=L$, then it requests that the neighbor nodes only take state $c_{k}=L-1$ or $c_{k}=L$, but at least one $L$-distance neighbor must be occupied. $\beta$ denotes the inverse temperature, and $\partial i$ denotes the neighbor nodes of node $i$. The partition function therefore only takes into account all the L-distance dominating set (DS).\\

RS mean field theories, such as the Bethe--Peierls approximation\cite{MMMA2009} and partition function expansion\cite{XJQZHJ2011,ZHJWC2012}, can solve the above spin glass model. These two theories obtain the same results, but the Bethe--Peierls approximation theory equation is easier to read; thus, we introduce the Bethe--Peierls approximation equation. We set cavity message $p_{i\rightarrow j}^{(c_{i},c_{j})}$ on each edge, and the message must satisfy 

\begin{equation}
p_{i\rightarrow j}^{(c_{i},c_{j})}=\frac{e^{-\beta \delta_{ c_{i}}^{0}}\prod\limits_{k\in\partial i\backslash j}\sum\limits_{c_{k}\in A}p_{k\rightarrow i}^{( c_{k}, c_{i})}-(1-\delta_{ c_{i}}^{0})(\delta_{ c_{j}}^{c_{i}}+\delta_{ c_{j}}^{c_{i}+1})\prod\limits_{k\in\partial i\backslash j}\sum\limits_{c_{k}\geq c_{i}}p_{k\rightarrow i}^{(c_{k},c_{i})}}{\sum\limits_{\acute{ c}_{i},\acute{ c}_{j}}e^{-\beta \delta_{\acute c_{i}}^{0}}\prod\limits_{k\in\partial i\backslash j}\sum\limits_{\acute c_{k}\in A}p_{k\rightarrow i}^{( \acute c_{k}, \acute c_{i})}-(1-\delta_{ \acute c_{i}}^{0})(\delta_{ \acute c_{j}}^{\acute c_{i}}+\delta_{ \acute c_{j}}^{\acute c_{i}+1})\prod\limits_{k\in\partial i\backslash j}\sum\limits_{\acute c_{k}\geq \acute c_{i}}p_{k\rightarrow i}^{(\acute c_{k},\acute c_{i})}},\end{equation}

which is called the BP equation, cavity message $p_{i\rightarrow j}^{(c_{i},c_{j})}$ represents the joint probability that node $i$ is in occupation state $c_{i}$ and its adjacent node $j$ is in occupation state $c_{j}$ when the constraint of node $j$ is not considered. Set $A$ represents the possible states of $c_{k}$. Marginal probability $p_{i}^{c}$ of node $i$ is expressed as
\begin{equation}
p_{i}^{c}=\frac{e^{-\beta \delta_{c}^{0}}\prod\limits_{j\in\partial i}\sum\limits_{c_{j}\in A}p_{j\rightarrow i}^{( c_{j}, c)}-(1-\delta_{ c}^{0})\prod\limits_{j\in\partial i}\sum\limits_{c_{j}\geq c}p_{j\rightarrow i}^{(c_{j},c)}}{\sum\limits_{c_{i}}e^{-\beta \delta_{c_{i}}^{0}}\prod\limits_{j\in\partial i}\sum\limits_{c_{j}\in A}p_{j\rightarrow i}^{(c_{j}, c_{i})}-(1-\delta_{ c_{i}}^{0})\prod\limits_{j\in\partial i}\sum\limits_{c_{j}\geq c_{i}}p_{j\rightarrow i}^{(c_{j},c_{i})}}.
\end{equation}

Messages $p_{j\rightarrow i}^{( c_{j}, c)}$ are converged messages, that is, the marginal probability is calculated after the BP equation converges. $p_{i}^{0}$ denotes the probability that node $i$ is covered, $p_{i}^{1}$ denotes the probability that node $i$ has at least one covered neighbor, $p_{i}^{2}$ denotes the probability that node $i$ has at least one covered 2-distance neighbor, and $p_{i}^{3}$ denotes the probability that node $i$ has at least one covered 3-distance neighbor, and so on.

Finally, the free energy can be calculated using mean field theory:
\begin{equation}
F_{0}=\sum_{i=1}^{N}F_{i}-\sum_{(i,j)=1}^{M}F_{(i,j)},
\end{equation}
where

\begin{equation}
F_{i}=-\frac{1}{\beta}\ln[\sum\limits_{c_{i}}e^{-\beta \delta_{c_{i}}^{0}}\prod\limits_{j\in\partial i}\sum\limits_{c_{j}\in A}p_{j\rightarrow i}^{(c_{j}, c_{i})}-(1-\delta_{ c_{i}}^{0})\prod\limits_{j\in\partial i}\sum\limits_{c_{j}\geq c_{i}}p_{j\rightarrow i}^{(c_{j},c_{i})}]
\end{equation}

\begin{equation}
F_{(i,j)}=-\frac{1}{\beta}\ln[\sum_{ c_{i}, c_{j}\in A}p_{i\rightarrow j}^{( c_{i}, c_{j})}p_{j\rightarrow i}^{( c_{j}, c_{i})}],
\end{equation}

where $F_{i}$ denotes the free energy of function node $i$ and $F_{(i,j)}$ denotes the free energy of edge $(i,j)$. The BP equation is iterated until it converges to one stable point, and then mean free energy $f\equiv F/N$ and energy density $e=1/N\sum_{i}p_{i}^{0}$ are calculated using (3) and (4). The entropy density is calculated as $s=\beta(e-f)$.

\begin{figure}[!htb]
  \centering
  \includegraphics[width=12.5cm,height=9cm]{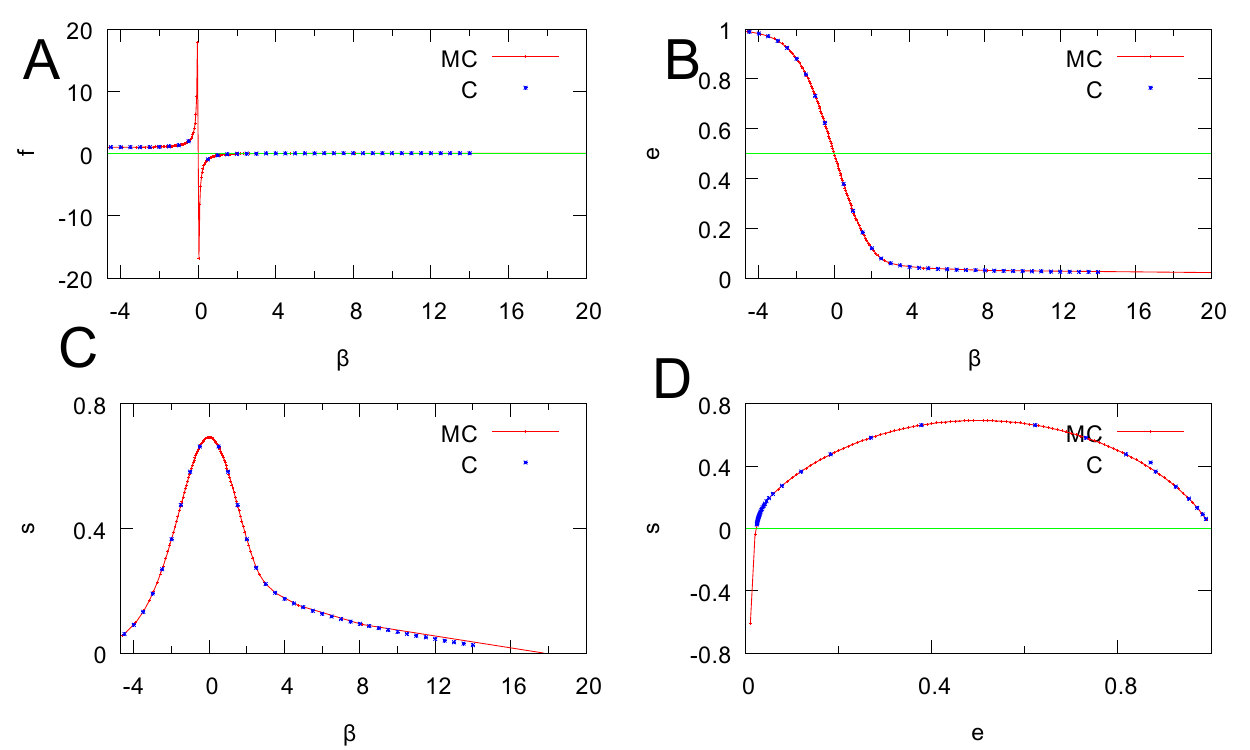}
  \caption{canonical BP and microcanonical BP results for the 2-distance MDS problem on the RR random graph with mean connectivity $c=10$ and $N=10000$. In the A,B,C graphs, the $x$-axis denotes the inverse temperature $\beta $ and the $y$-axis denotes the thermodynamic quantities. In the graph D, the $x$-axis denotes the energy density and the $y$-axis denotes the entropy density.}
\end{figure}

The canonical BP equation not always converge to a fixed point at ground state energy for any $L$-distance MDS so that we can't exactly predict the ground state energy. For example,  it converges to a fixed point at ground state when node degree only in the range from 3 to 9 for $L=2$ and from 3 to 4 for $L=3$ and empty for $L=4$. Although canonical Replica Symmetry theory can find a fixed point at lower $\beta$ for any $L$-distance MDS, it can't find the ground state energy. Fig2 and Fig3 indicate that canonical BP equation can't converge to a stable point while microcanonical BP equation still converge to a stable point when $\beta$ bigger than a threshold value, we suppose that solution space has a phase transition at this threshold value. Fig4 and Fig5 indicate that the canonical RS theory outperform the microcanonical RS theory, but they both can not find the ground state energy. Table1 gives the ground state energy of $L\leq 3$ distance MDS by microcanonical BP equation. We find that the microcanonical process results agreement with canonical ensemble process where canonical BP equation can predict the ground state energy. With the increase of mean node degree and $L$, the convergence speed of the microcanonical BP equation becomes slower and requires a larger damping factor.
\begin{figure}[!htb]
  \centering
  \includegraphics[width=12.5cm,height=9cm]{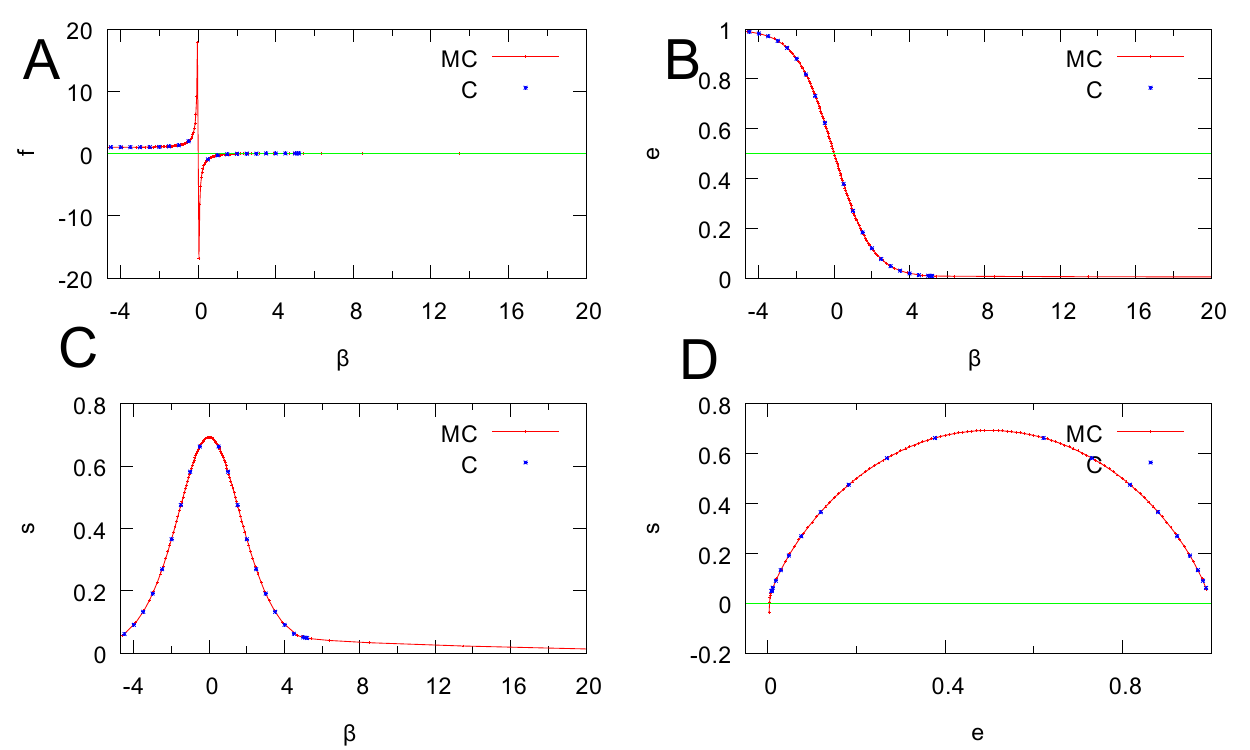}
  \caption{canonical BP and microcanonical BP results for the 3-distance MDS problem on the RR random graph with mean connectivity $c=10$ and $N=10000$. In the A,B,C graphs, the $x$-axis denotes the inverse temperature $\beta $ and the $y$-axis denotes the thermodynamic quantities. In the graph D, the $x$-axis denotes the energy density and the $y$-axis denotes the entropy density.}
\end{figure}
\begin{table}[!hbp]
\caption{The ground state energy density for RR random graph when $L$ from1 to 4.}
\begin{tabular}{p{0.8cm}p{0.8cm}p{0.8cm}p{0.8cm}p{0.8cm}p{0.8cm}p{0.8cm}p{0.8cm}p{0.8cm}p{0.8cm}}
\hline
C & 3 & 4 & 5 & 6 & 7 & 8 & 9 &10 & 11\\
\hline
$E_{min}^{L=1}\approx$ & 0.264 & 0.223 & 0.195 & 0.173 & 0.157 & 0.144 & 0.133 & 0.123  & 0.117 \\
\hline
C & 12 & 13 & 14 & 15 & 16 & 17 & 18 & 19 & 20\\
\hline
$E_{min}^{L=1}\approx$ & 0.110 & 0.104 & 0.100 & 0.095 & 0.090 & 0.087 & 0.083 & 0.080  & 0.077 \\
\hline
C & 3 & 4 & 5 & 6 & 7 & 8 & 9 &10 & 11\\
\hline
$E_{min}^{L=2}\approx$ & 0.11 & 0.08 & 0.06 & 0.047 & 0.038 & 0.03 & 0.027 & 0.022  & 0.019 \\
\hline
C & 12 & 13 & 14 & 15 & 16 & 17 & 18 & 19 & 20\\
\hline
$E_{min}^{L=2}\approx$ & 0.017 & 0.015 & 0.013 & 0.012 & 0.011 & 0.01 & 0.009 & 0.008  & 0.008 \\
\hline
C & 3 & 4 & 5 & 6 & 7 & 8 & 9 &10 & 11\\
\hline
$E_{min}^{L=3}\approx$ & 0.06 & 0.033 & 0.02 & 0.014 & 0.009 & 0.007 & 0.005 & 0.004  & 0.003 \\
\hline
\end{tabular}
\end{table}

\begin{figure}[!htb]
  \centering
  \includegraphics[width=12.5cm,height=9cm]{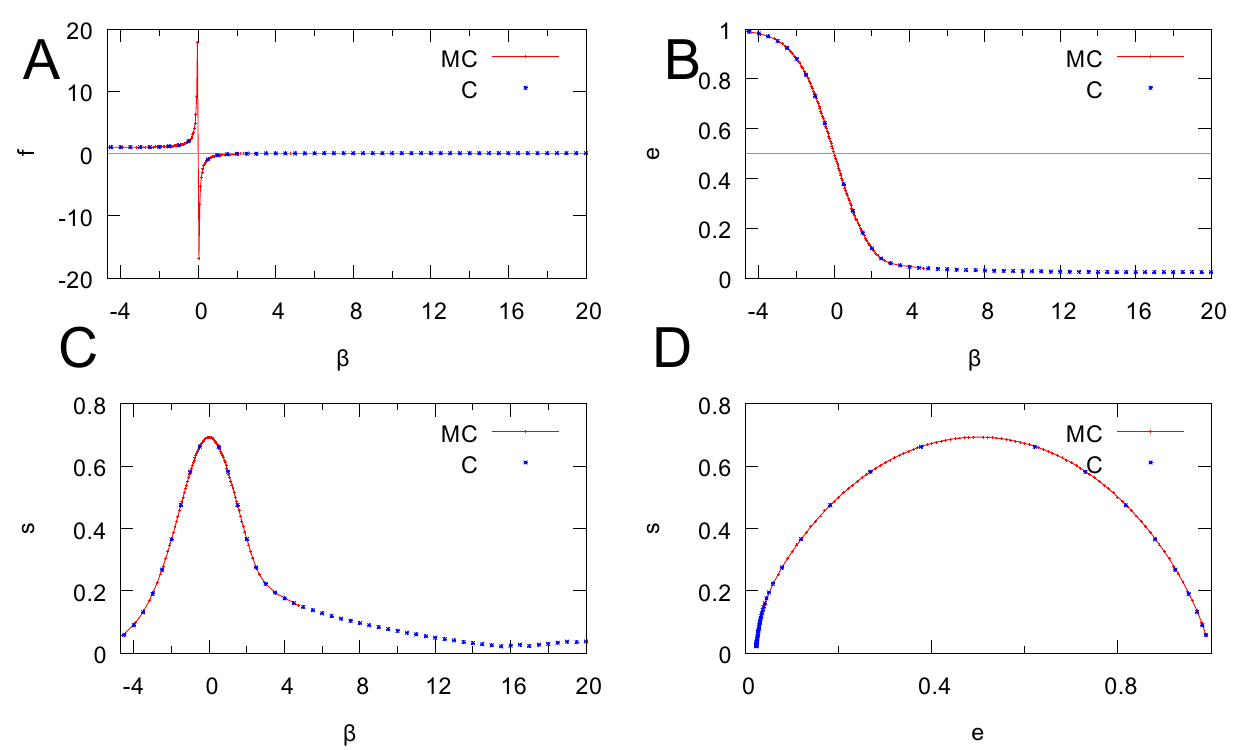}
  \caption{canonical RS and microcanonical RS results for the 2-distance MDS problem on the RR random graph with mean connectivity $c=10$ and $N=10000$. In the A,B,C graphs, the $x$-axis denotes the inverse temperature $\beta $ and the $y$-axis denotes the thermodynamic quantities. In the graph D, the $x$-axis denotes the energy density and the $y$-axis denotes the entropy density.}
\end{figure}

\begin{figure}[!htb]
  \centering
  \includegraphics[width=12.5cm,height=9cm]{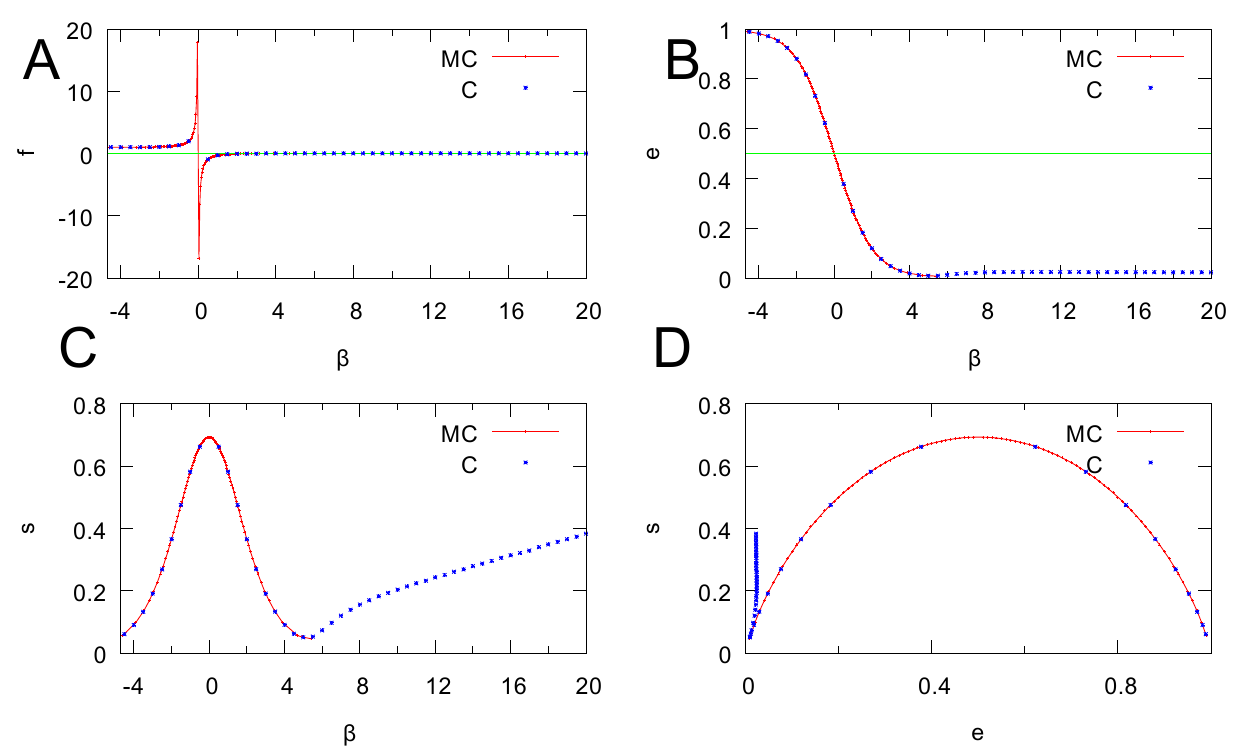}
  \caption{canonical RS and microcanonical RS results for the 3-distance MDS problem on the RR random graph with mean connectivity $c=10$ and $N=10000$. In the A,B,C graphs, the $x$-axis denotes the inverse temperature $\beta $ and the $y$-axis denotes the thermodynamic quantities. In the graph D, the $x$-axis denotes the energy density and the $y$-axis denotes the entropy density.}
\end{figure}

\begin{figure}[!htb]
  \centering
  \includegraphics[width=12.5cm,height=9cm]{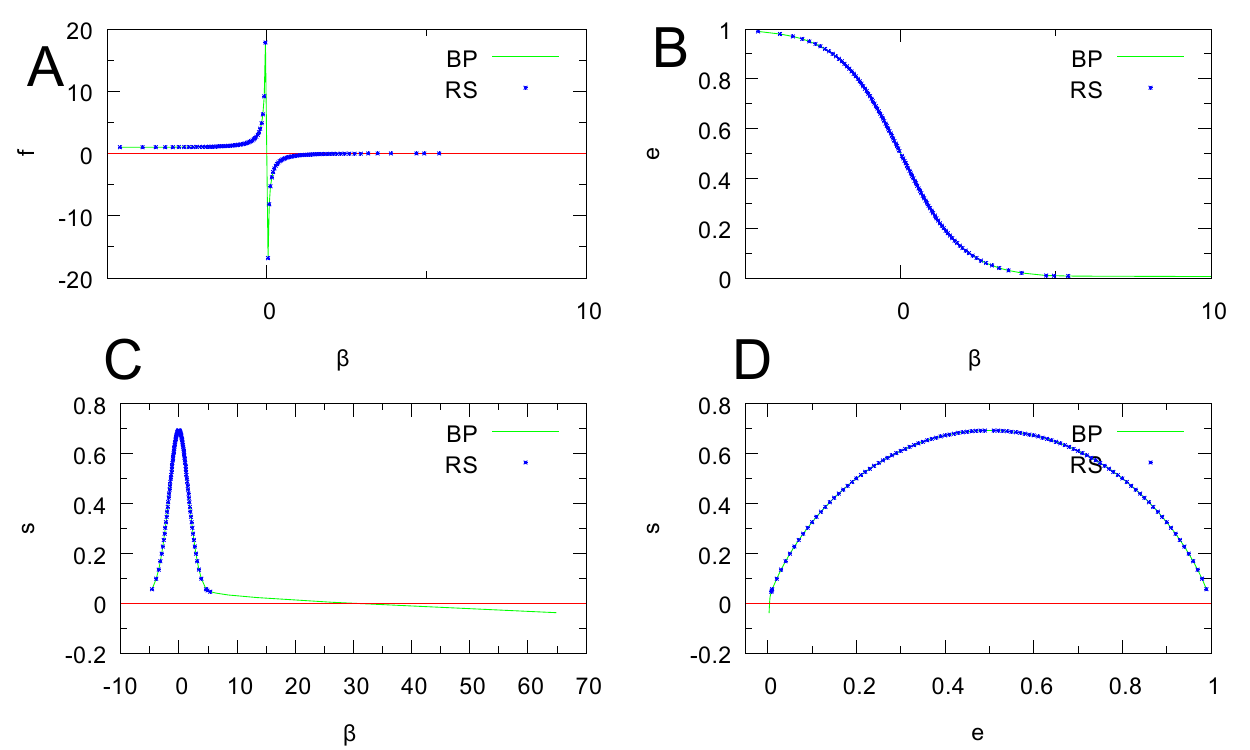}
  \caption{RS and BP results for the 3DMDS problem on the RR random graph with mean connectivity $c=10$ and $N=10^{4}$ using microcanonical BP and microcanonical population dynamics. In the A,B,C graphs, the $x$-axis denotes the inverse temperature $\beta $ and the $y$-axis denotes the thermodynamic quantities. In the graph D, the $x$-axis denotes the energy density and the $y$-axis denotes the entropy density.}
\end{figure}
\begin{table}[!hbp]
\caption{The energy density of negative temperature phase transition for RR random graph when $L$ from1 to 4.}
\begin{tabular}{p{0.8cm}p{0.8cm}p{0.8cm}p{0.8cm}p{0.8cm}p{0.8cm}p{0.8cm}p{0.8cm}p{0.8cm}p{0.8cm}p{0.8cm}}
\hline
C & 1 & 2 & 3 & 4 & 5 & 6 & 7 & 8 & 9 &10\\
\hline
$E_{\beta^{-}}^{L=1}\approx$ & 0.67 & 0.62 & 0.57 & 0.55 & 0.53 & 0.52 & 0.513 & 0.51 & 0.505 & 0.5025 \\
\hline
C & 11 & 12 & 13 & 14 & 15 & 16 & 17 & 18 & 19 & 20\\
\hline
$E_{\beta^{-}}^{L=1}\approx$ & 0.50 & 0.50 & 0.50 & 0.50 & 0.50 & 0.50 & 0.50 & 0.50  & 0.50 & 0.50 \\
\hline
C & 1 & 2 & 3 & 4 & 5 & 6 & 7 & 8 & 9 &10\\
\hline
$E_{\beta^{-}}^{L=2}\approx$ & 0.67 & 0.54 & 0.504 & 0.50 & 0.50 & 0.50 & 0.50 & 0.50  & 0.50 & 0.50 \\
\hline
C & 1 & 2 & 3 & 4 & 5 & 6 & 7 & 8 & 9 & 10\\
\hline
$E_{\beta^{-}}^{L=3}\approx$ & 0.67 & 0.53 & 0.50 & 0.50 & 0.50 & 0.50 & 0.50 & 0.50  & 0.50 & 0.50 \\
\hline
C & 1 & 2 & 3 & 4 & 5 & 6 & 7 & 8 & 9 &10\\
\hline
$E_{\beta^{-}}^{L=4}\approx$ & 0.67 & 0.51 & 0.50 & 0.50 & 0.50 & 0.50 & 0.50 & 0.50  & 0.50 & 0.50 \\
\hline
\end{tabular}
\end{table}

We use microcanonical BP process to predict ground state energy of $10^4$ node single graph instance, but the ground state prediction still valuable for any graph instance.
From the Figure2-Figure 5 we can see that the inverse temperature changes from negative to positive when the energy density drops to a certain threshold. In Table 2, we show the relationship between the energy threshold value and the mean node connectivity in different L-distance. 
\section{Belief propagation decimation algorithm and greedy algorithm}
In this paper, we use two algorithms to determine the solution of the given graph: the greedy algorithm and Canonical BPD algorithm. The greedy algorithm very fast, but it does not guarantee good results such as Canonical BPD. The Canonical BPD algorithm is not as fast as the greedy algorithm, but it always provides a good estimation for the L-distance MDS problem.  
\subsection{Belief Propagation Decimation}

If node $i$ is unobserved, then output message $p_{i\rightarrow j}$ on the link $(i,j)$ between node $j$ and node $i$ is updated according to Eq.(2).By contrast, if node $i$ is empty but observed (it is not occupied, but it has at least one occupied l-distance neighbor node), we record the observed state as $c^{*}$, this node then presents no restriction to the occupation states of all its unoccupied l-distance neighbors. For such a node i, output message $p_{i\rightarrow j}$ on the link $(i,j)$ is then updated according to 
\begin{equation}
p_{i\rightarrow j}^{(c_{i},c_{j})}=\frac{e^{-\beta \delta_{ c_{i}}^{0}}\Theta(c^{*}-c_{i})[\prod\limits_{k\in\partial i\backslash j}\sum\limits_{c_{k}\in A}p_{k\rightarrow i}^{( c_{k}, c_{i})}-R(c_{i},c_{j})\prod\limits_{k\in\partial i\backslash j}\sum\limits_{c_{k}\geq c_{i}}p_{k\rightarrow i}^{(c_{k},c_{i})}]}{\sum\limits_{\acute{ c}_{i},\acute{ c}_{j}}e^{-\beta \delta_{\acute c_{i}}^{0}}\Theta(c^{*}-\acute c_{i})[\prod\limits_{k\in\partial i\backslash j}\sum\limits_{\acute c_{k}\in A}p_{k\rightarrow i}^{( \acute c_{k}, \acute c_{i})}-R(\acute c_{i},\acute c_{j})\prod\limits_{k\in\partial i\backslash j}\sum\limits_{\acute c_{k}\geq \acute c_{i}}p_{k\rightarrow i}^{(\acute c_{k},\acute c_{i})}]}.
\end{equation}
\begin{equation}
R(c_{i},c_{j})=(1-\delta_{ c_{i}}^{0}-\delta_{ c_{i}}^{c^{*}})(\delta_{ c_{j}}^{c_{i}}+\delta_{ c_{j}}^{c_{i}+1})
\end{equation}

 in there the function $\Theta(x)=1$ if $x\geq 0$ and $\Theta(x)=0$ otherwise. For node $i (c_{i}=c^{*})$, if at least one neighbor node $j$ takes state $c_{j}=c^{*}-1$, then it sends a message to node $i$ as $\sum\limits_{c}p_{j\rightarrow i}^{(c^{*}, c)}=0$.  It leads $\sum\limits_{c}p_{j\rightarrow i}^{(c, c^{*})}=p_{j\rightarrow i}^{(c_{*}-1, c_{*})}$, so the constraints of node $i$ to all the other neighbor nodes are automatically removed.  The marginal probability is calculated by 

\begin{equation}
p_{i}^{c}=\frac{e^{-\beta \delta_{c}^{0}}\Theta(c^{*}-c)[\prod\limits_{j\in\partial i}\sum\limits_{c_{j}\in A}p_{j\rightarrow i}^{( c_{j}, c)}-(1-\delta_{ c}^{0}-\delta_{ c}^{c^{*}})\prod\limits_{j\in\partial i}\sum\limits_{c_{j}\geq c}p_{j\rightarrow i}^{(c_{j},c)}]}{\sum\limits_{c_{i}}e^{-\beta \delta_{c_{i}}^{0}}\Theta(c^{*}-c)[\prod\limits_{j\in\partial i}\sum\limits_{c_{j}\in A}p_{j\rightarrow i}^{(c_{j}, c_{i})}-(1-\delta_{ c_{i}}^{0}-\delta_{ c_{i}}^{c^{*}})\prod\limits_{j\in\partial i}\sum\limits_{c_{j}\geq c_{i}}p_{j\rightarrow i}^{(c_{j},c_{i})}]}.
\end{equation}

in this work we construct a solution for the L-distance $(2<L<7)$ MDS problem by BPD algorithm.
We implement the BPD algorithm as follows:\\
(1) Input network $W$, set all the nodes to be unobserved, and set all the cavity messages $p_{i\rightarrow j}^{(c_{i},c_{j})}$ to be uniform messages. Set inverse temperature $\beta$ to be sufficiently large (depending on the at most convergence inverse temperature). Then iterate the BP equation using Eq.(2) until it converges to one stable point. Finally, compute the occupation probability of each node $i$ using Eq.(3).  \\
(2) Cover one of unfixed nodes that have the highest covering probabilities.\\
(3) Update the state of all the uncovered nodes as follows: if node $i$ is uncovered and has at least one neighbor that takes state $c_{i}=0$, then it takes state $c_{i}=1$, and if node $i$ is uncovered and has at least one neighbor that takes state $c_{i}=1$, then it takes state $c_{i}=2$. If node $i$ is uncovered and has at least one neighbor that takes state $c_{i}=2$, then it takes state $c_{i}=3$, and so on.\\
(4) Fix the observed node's state, that is, if all the neighbor nodes of observed node $c_{i}=1$ are covered or in the state $c_{j}=1$, then fix the state of node $i$ to $c_{i}=1$. \\
(5) If network $W$ still contains unobserved nodes, then perform the BP equation using Eqs.(2) or (7). Calculate the marginal probability using Eqs.(4) or (9), depending on the state of node $i$. Repeat operations (2)--(4) until all nodes are observed. 
\subsection{Greedy}
We develop a very simple greedy algorithm in the literature to solve the L-distance MDS problem approximately, which is based on the concept of the node's impact. The impact of unoccupied node $i$ equals the number of nodes that will be observed in L-distance by occupying $i$. Starting from input network W with all the nodes unobserved, the greedy algorithm uniformly selects at random node i from the subset of nodes with the highest general impact and fixes its occupation state to $c_{i}$ = 0. Then all the neighbor nodes in L-distance of node $i$ are observed. If there are still unobserved nodes in the network, then the impact value for each of the unoccupied nodes is updated and the greedy occupying process is repeated until all the nodes are observed. This pure greedy algorithm is very easy to implement and very fast. We found that it typically reaches a true L-distance MDS when the input network contains more edges.

\begin{figure}[htb]
  \centering
  \includegraphics[width=12cm,height=7cm]{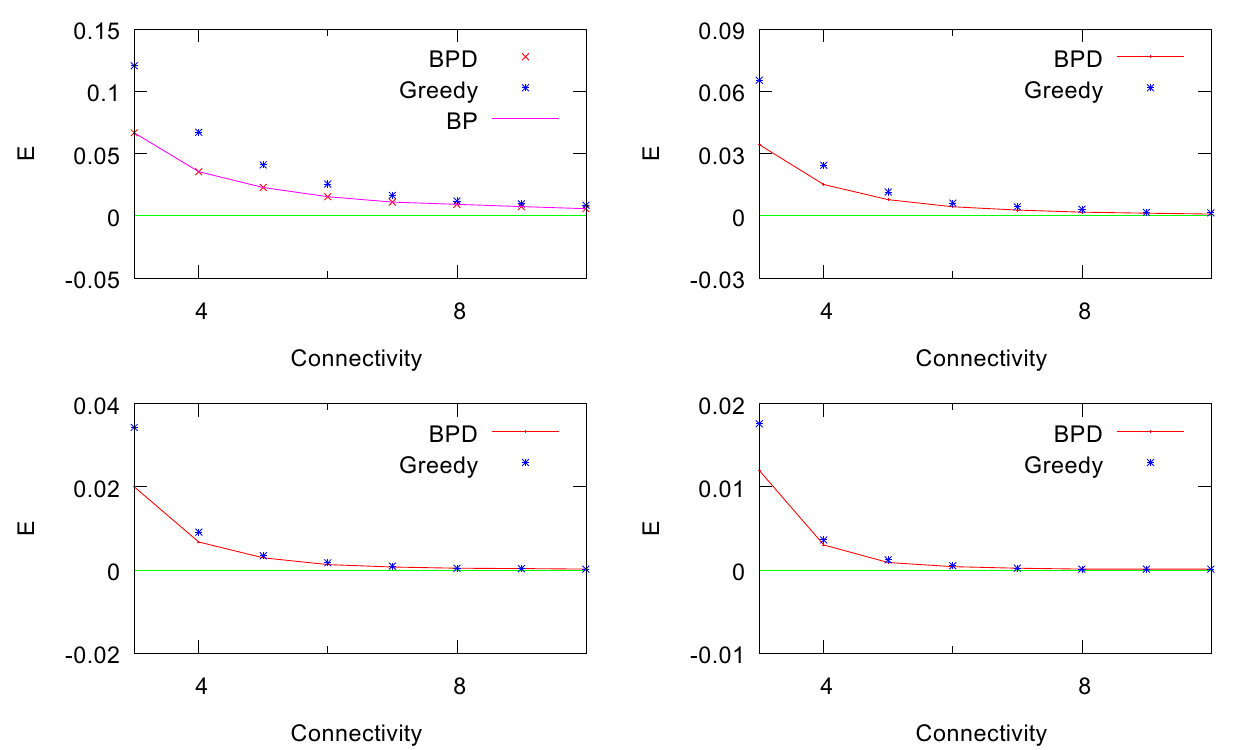}
  \caption{BPD,Greedy and microcanonical BP results for the 3-distance MDS problem on the 16 RR graphs with the size of $N=10^{4}$ nodes. The $x$-axis denotes the mean connectivity and the $y$-axis denotes the energy density. From the upper left graph, the order is L=3,L=4,L=5 and L=6. }
\end{figure}
\begin{figure}[htb]
  \centering
  \includegraphics[width=12cm,height=7cm]{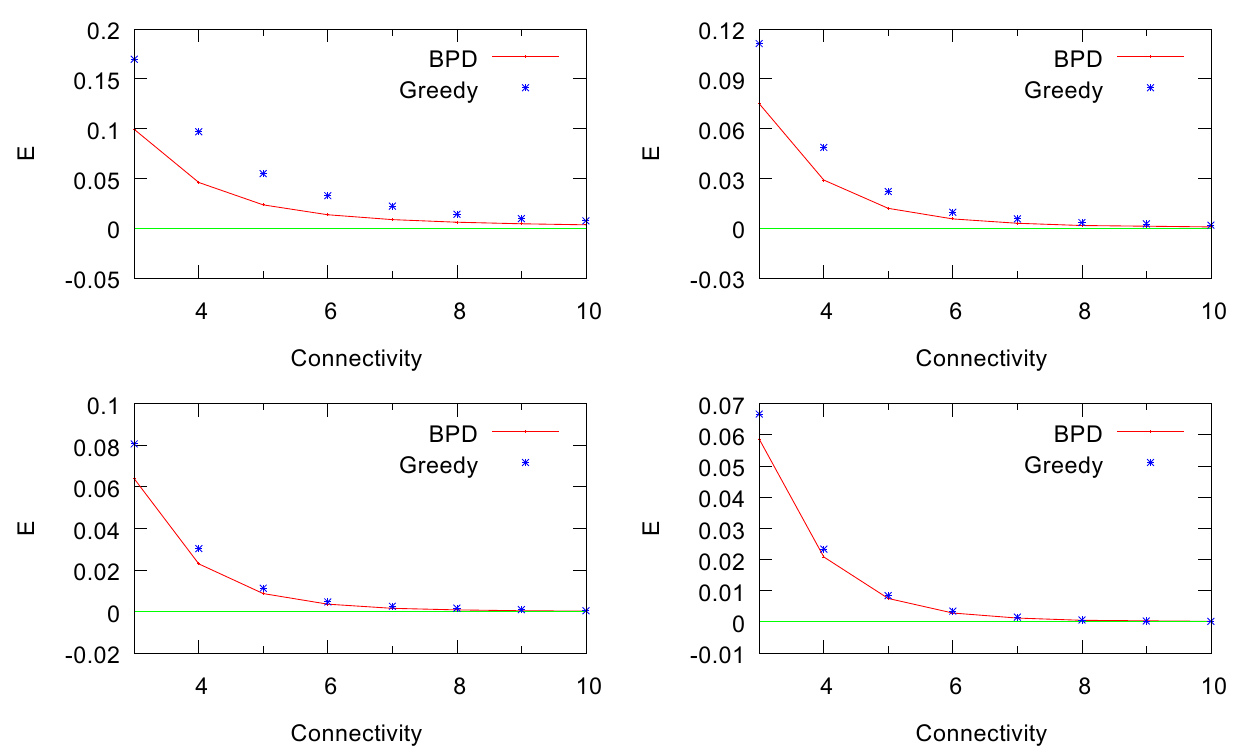}
  \caption{BPD,Greedy and microcanonical BP results for the 3-distance MDS problem on the 16 ER random graphs with the size of $N=10^{4}$ nodes. The $x$-axis denotes the mean connectivity and the $y$-axis denotes the energy density. From the upper left graph, the order is L=3,L=4,L=5 and L=6. }
\end{figure}
\begin{figure}[htb]
  \centering
  \includegraphics[width=12cm,height=7cm]{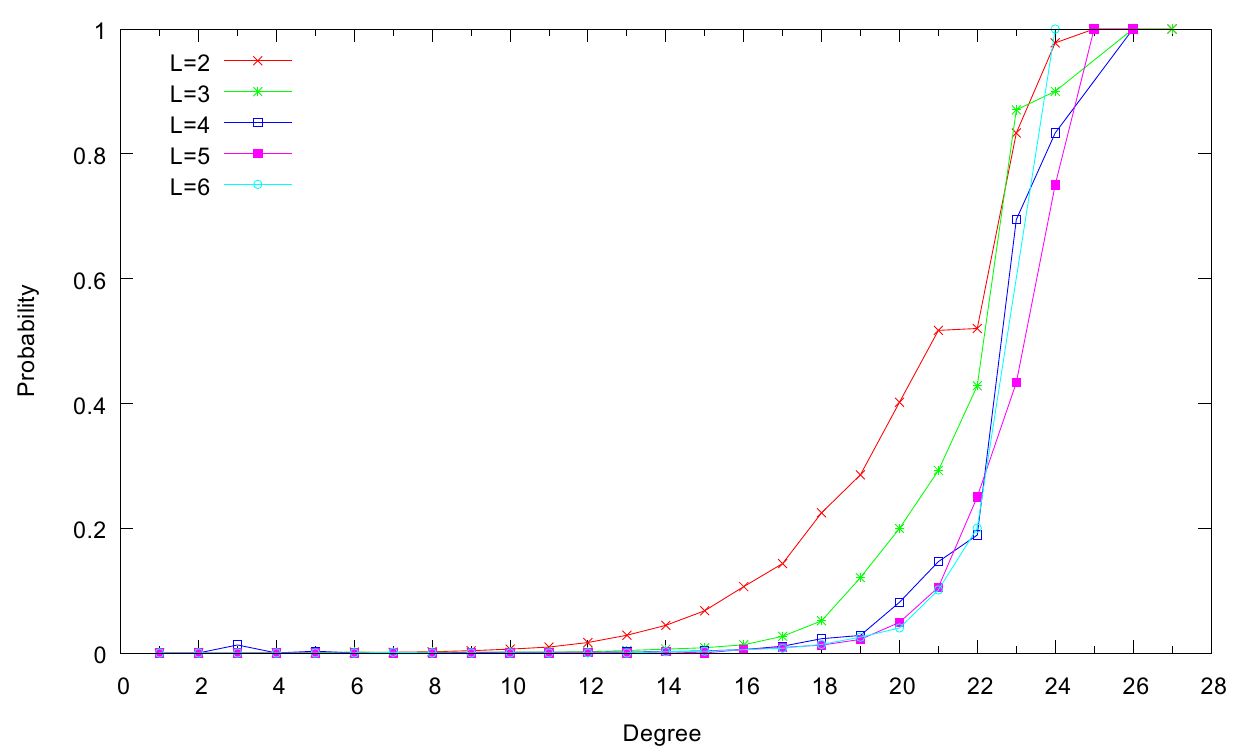}
  \caption{The occupied probability of a vertex degree for the L-distance $(1<L<7)$ MDS problem on the ER random graph with the size of $N=10^{4}$ nodes and average degree 10. The results obtained by BPD algorithm on 16 random graph instances.The $x$-axis denotes the degree and the $y$-axis denotes the probability.}
\end{figure}

The results of the BPD algorithm for the RR and ER random networks are compared with the results of the greedy algorithm in Figs. 7 and 8. The BPD algorithm outperformed the greedy algorithm, and  provided results that were always very close to those of the microcanonical BP equation prediction. We can see that the Greedy results very close to the results of the BPD results when the $L$ and degree are more large. In a ER random graph with abundant connections, the vertices with large degree are more likely belonging to the L-distance MDS, and the vertices connecting with a few neighbors tend to be observed by their high degree neighbors.This phenomenon is observed by Fig.9, where we present the occupying probability of different vertex degree in ER random graph. The occupying probability of almost vertex degrees decrease with the $L$ except few high degrees. All the vertex of a ER random network with mean degree 10 can be observed by occupying one or two vertex when $L=6$, in this case we can not calculate the occupying probability of different degrees by averaging over all graphs, we only average over occupied degrees on different graphs.       
\section{Solving the Belief-Propagation equation}
We use two different processes to solve the BP equation, namely Canonical ensemble process and Microcanonical ensemble process. In Canonical ensemble process, particle number $N$, volume $V$ and temperature $T$ are invariant. In each step, we look for the corresponding equilibrium states to these invariants of $N,V$ and $T$.  In Microcanonical ensemble process, particle number $N$, volume $V$ and energy $E$ are invariant. In this process, we look for the corresponding equilibrium states to these invariants of $N,V$ and $E$ at each step.
\subsection{Canonical ensemble process}
In the Canonical BP process, the $N$ and $V$ are always invariants, so we just need to fix the temperature. At a given fixed value of $\beta$, we iterate the BP equation on a single graph $G$ to obtain a fixed-point solution. At each elemental iteration process a vertex $i$ is chosen from all $N$ vertices of the graph, and the cavity probability distributions $p_{i\rightarrow j}^{c_{i},c_{j}}$ on the edges between $i$ and all its nearest neighbors $j$ are updated according to Eq. (2). When $G$ is a RR graph we experience that this BP evolution converges to a fixed point within about $100N$ elemental updates, and this fixed point is uniform in that the cavity probability distributions are identical for all the graph edges.\\
To get ensemble-averaged results for random graphs characterized by certain vertex degree profile, we also perform population dynamics simulations based on Eq. (2). In the case of the RR graph ensemble, we first construct a long array of cavity probability distributions $p_{i\rightarrow j}^{c_{i},c_{j}}$; then we repeatedly update it by (1) drawing $K-1$ cavity distributions uniformly at random from this array as inputs to Eq. (2) to generate a new cavity distribution, and (2) replace an old cavity distribution in the
array (chosen uniformly at random) by this new cavity distribution. This population dynamics also drives the population of cavity probability distributions to the uniform population (all the elements being identical) for the RR graph ensemble. The ensemble-averaged and single graph BP results therefore are in complete agreement. \\
\subsection{Microcanonical ensemble process}
In the Microcanonical BP process, the $N$ and $V$ are always invariants, so we just need to fix the energy density $e$.To perform BP iteration at fixed energy density $e$, we need to slightly modify Eq. (2) as follows(L=3 example)
\begin{equation}
\omega_{i\rightarrow j}^{0,0}=\omega_{i\rightarrow j}^{0,1}=\prod\limits_{k\in\partial i\backslash j}(p_{k\rightarrow i}^{0,0}+p_{k\rightarrow i}^{1,0})
\end{equation}
\begin{equation}
\omega_{i\rightarrow j}^{1,0}=\prod\limits_{k\in\partial i\backslash j}\sum\limits_{c_{k}\in A}p_{k\rightarrow i}^{c_{k},1}
\end{equation}
\begin{equation}
\omega_{i\rightarrow j}^{1,1}=\omega_{i\rightarrow j}^{1,2}=\prod\limits_{k\in\partial i\backslash j}\sum\limits_{c_{k}\in A}p_{k\rightarrow i}^{c_{k},1}-\prod\limits_{k\in \partial i\backslash j}\sum\limits_{c_{k}\in A\backslash 0}p_{k\rightarrow i}^{c_{k},1}
\end{equation}
\begin{equation}
\omega_{i\rightarrow j}^{2,1}=\prod\limits_{k\in\partial i\backslash j}\sum\limits_{c_{k}\in A}p_{k\rightarrow i}^{c_{k},2}
\end{equation}
\begin{equation}
\omega_{i\rightarrow j}^{2,2}=\omega_{i\rightarrow j}^{2,3}=\prod\limits_{k\in\partial i\backslash j}\sum\limits_{c_{k}\in A}p_{k\rightarrow i}^{c_{k},2}-\prod\limits_{k\in \partial i\backslash j}\sum\limits_{c_{k}\in A\backslash 1}p_{k\rightarrow i}^{c_{k},2}
\end{equation}
\begin{equation}
\omega_{i\rightarrow j}^{3,2}=\prod\limits_{k\in\partial i\backslash j}\sum\limits_{c_{k}\in A}p_{k\rightarrow i}^{c_{k},3}
\end{equation}
\begin{equation}
\omega_{i\rightarrow j}^{3,3}=\prod\limits_{k\in\partial i\backslash j}\sum\limits_{c_{k}\in A}p_{k\rightarrow i}^{c_{k},3}-\prod\limits_{k\in \partial i\backslash j}p_{k\rightarrow i}^{3,3}
\end{equation}
where $\omega_{i\rightarrow j}^{0,0},\omega_{i\rightarrow j}^{0,1},\omega_{i\rightarrow j}^{1,0},\omega_{i\rightarrow j}^{1,1},\omega_{i\rightarrow j}^{1,2},\omega_{i\rightarrow j}^{2,1},\omega_{i\rightarrow j}^{2,2},\omega_{i\rightarrow j}^{2,3},\omega_{i\rightarrow j}^{3,2}$ and $\omega_{i\rightarrow j}^{3,3}$ are ten auxiliary weight messages from vertex $i$ to its nearest neighbor $j$. We denote these ten real quantities collectively as $\bm{\omega}_{i\rightarrow j}$.
Similarly, we denote the marginal weights $\bm{\omega_{i}}\equiv(\omega_{i}^{0},\omega_{i}^{1},\omega_{i}^{2},\omega_{i}^{3})$
of vertex $i$ as
\begin{equation}
\omega_{i}^{0}=\prod\limits_{j\in\partial i}(p_{j\rightarrow i}^{0,0}+p_{j\rightarrow i}^{1,0})
\end{equation}
\begin{equation}
\omega_{i}^{1}=\prod\limits_{j\in\partial i}\sum\limits_{c_{j}\in A}p_{j\rightarrow i}^{c_{j},1}-\prod\limits_{j\in\partial i}\sum\limits_{c_{j}\in A\backslash 0}p_{j\rightarrow i}^{c_{j},1}
\end{equation}
\begin{equation}
\omega_{i}^{2}=\prod\limits_{j\in\partial i}\sum\limits_{c_{j}\in A}p_{j\rightarrow i}^{c_{j},2}-\prod\limits_{j\in\partial i}\sum\limits_{c_{j}\in A\backslash 1}p_{j\rightarrow i}^{c_{j},2}
\end{equation}
\begin{equation}
\omega_{i}^{3}=\prod\limits_{j\in\partial i}\sum\limits_{c_{j}\in A}p_{j\rightarrow i}^{c_{j},3}-\prod\limits_{j\in\partial i}p_{j\rightarrow i}^{3,3}
\end{equation}
In each BP iteration the following actions are taken: (1) we update the output messages $\bm{\omega_{i\rightarrow j}}$ and $\bm{\omega_{j\rightarrow i}}$ for each pair of edges $(i,j)$ of the graph according to Eq.(10-16), and the marginal weights $\omega_{i}$ for all the vertices $i$ according to Eq.(17-20); (2) and determine the value of the inverse temperature $\beta$ as the root of the following equation
\begin{equation}
\bar{E}=\frac{e^{-\beta}\sum\limits_{i=1}^{N}\omega_{i}^{0}}{e^{-\beta}\sum\limits_{i=1}^{N}\omega_{i}^{0}+\sum\limits_{i=1}^{N}(\omega_{i}^{1}+\omega_{i}^{2}+\omega_{i}^{3})}
\end{equation}
this formula guarantee the BP equation can find the ground state energy, so we can write the inverse temperature as
\begin{equation}
\beta=log[(1-\bar{E})\times\sum\limits_{i=1}^{N}\omega_{i}^{0}]-log[\bar{E}\times\sum\limits_{i=1}^{N}(\omega_{i}^{1}+\omega_{i}^{2}+\omega_{i}^{3})]
\end{equation}

\section{Discussion}
In this paper, we proposed two heuristic algorithms (a greedy-impact local algorithm and a BPD message-passing algorithm) and presented an RS mean field theory, developing canonical and microcanonical thermodynamical process for l-distance dominating set problem. We found that the results of canonical BP algorithm agreement with the results of RS mean field theory. In the 2-distance MDS problem, the entropy function $S(E)$ has an phase transition in the RR network when the connectivity is only from 3 to 9. In the 3-distance MDS problem, the entropy function $S(E)$ has an phase transition in the RR network when the connectivity is only from 3 to 4. We found that the results of microcanonical BP algorithm doesn't agreement with the results of RS mean field theory. In the 2-distance MDS problem, the entropy function $S(E)$ has an phase transition in the RR network when the connectivity is big enough. In the 3-distance MDS problem, the entropy function $S(E)$ has an phase transition in the RR network when the connectivity is from 3 to 11. Our numerical results shown in Figs. 7 and 8 suggest that the BPD algorithm always outperform the greedy heuristic algorithm.\\
A great deal of theoretical work remains to be studied. A direct extension of our work is to consider the l-distance MDS problem of the directed network. We will work on the directed l-distance MDS problem as soon as possible. A more challenging and common problem is the one step replica symmetry breaking theory of the l-distance MDS problem\cite{HYJSM2017}. We will use one step replica symmetry breaking theory to study both the canonical and microcanonical l-distance MDS problem. We still consider the Multi distance MDS problem as soon as possible, we think that this problem has very wide application in other field.
\section{$\hspace{2mm}$ Acknowledgement}
This research was supported by the doctoral startup fund of Xinjiang University of China (grant number 208-61357) and partially supported by the National Natural Science Foundation of China (grant numbers 11765021,11705279 and 61662078). 

\end{document}